\documentstyle[preprint,revtex]{aps}
\begin{document}
\draft
\preprint{IFT-P.023}
\preprint{IFUSP/P-997}
\preprint{July 1992}
\begin{title}
Three generations mixing and $\tau$ decay puzzle
\end{title}
\author{C. O. Escobar$^{a}$, O.L.G. Peres$^{b}$, V. Pleitez$^{b}$
\\and \\R. Zukanovich Funchal$^{a}$ }
\begin{instit}
$^{a}$ Instituto de F\'\i sica da Universidade de S\~ao Paulo\\
01498-970 C.P. 20516--S\~ao Paulo, SP\\
Brazil\\
$^{b}$Instituto de F\'\i sica Te\'orica \\
Universidade Estadual Paulista \\
Rua Pamplona, 145 \\
01405-900--S\~ao Paulo, SP \\
Brazil
\end{instit}
\begin{abstract}
We consider the possibility that the $\tau$ decay puzzle is a
consequence of the Kobayashi-Maskawa mixing in the leptonic sector.
\end{abstract}
The physics of the $\tau$ lepton will provide in the near future
evidence concerning the question if this lepton, with its
neutrino partner, is a sequential lepton or not. So far, as was
stressed in Ref.~\cite{eppz}, all experiments are internally consistent
with the Standard Model. Notwithstanding, it is well known that the
accuracy of the $\tau$ data are still poor and it should be possible that
new physics will come up when the proposed $\tau$-Charm factory gives new
and more accurate data about $\tau$ decays and properties~\cite{p}.

For example, new data could confirm the so called ``$\tau$ decay
puzzle''~\cite{m}, which is a difference between the measured
world average~\cite{pdg} $\tau$ lifetime, $\tau_\tau$ and the
theoretically expected value within two standard deviations.
Explicitly~\cite{hp}
\begin{equation}
\tau_{\tau}^{exp}-\tau^{th}_\tau=(0.16\pm0.09)\times10^{-13}\,s.
\label{1}
\end{equation}
although this discrepancy is not yet statistically significant, it
can be translated into discrepancies in particular branching ratios,
implying that the expected leptonic branching ratios are about
$2.3\sigma$ higher than the average measurements~\cite{m}. The
branching ratio conflicts suggest that a shift in $\tau_\tau$ and/or
$\tau$ mass, $m_\tau$, should occur when more precise measurements
become available. In fact, preliminary results from BES in Beijing
point to a down shift of $2\sigma$ in $m_\tau$ in relation to the
world average value~\cite{d}. Notwithstanding, in order to solve the
discrepancy a down shift of $6.4\sigma$ is required as pointed out in
Ref.~\cite{m}.

Other possibility is that no such a shift on $\tau_\tau$ or $m_\tau$
is needed but the problem with the branching ratios would still
exist. This would imply a new physics. The relationship of Eq.~(\ref{1})
with a possible deviation from universality can be see as follows. It
is well known that the decay diagram of the muon into electron being
similar to that of the $\tau$ decay into electron or muon. For
example, implies the following relationship
\begin{equation}
\left(\frac{G_\tau}{G_\mu}\right)^2=\left(\frac{\tau_\mu}
{\tau_\tau}\right)
\left(\frac{m_\mu}{m_\tau}\right)^5B.R.(\tau\to e\bar\nu\nu),
\label{2}
\end{equation}
where $G_\tau$ and $G_\mu$ are the coupling constant of the $\tau$ and
$\mu$ to the charged weak current respectively. Assuming $e\mu$
universality an average leptonic branching ratio can be defined as in
Ref.~\cite{d}:
\begin{equation}
\langle
B^\tau_l\rangle=\frac{B^\tau_e+\frac{B^\tau_\mu}{0.973}}{2}=17.88\pm0.26\,\%.
\label{3}
\end{equation}
where the factor $0.973$ is due to the mass of the muon.

Using the world average value for the branching ratio Eq.~(\ref{3}) in
Eq.~(\ref{2}) it follows
that $G_\tau/G_\mu=0.975\pm0.010$ using non-LEP and LEP data or
$G_\tau/G_\mu=0.985\pm0.0009$  using the BEPC value for the $\tau$
mass~\cite{d}. Of course in the Standard Model universality implies
$G_\tau=G_\mu$.

There have been some speculations about this possible deviation from the
Standard Model. Some examples, usually discussed in the literature, of
the new physics needed to solve the puzzle are:
\begin{enumerate}
\item the introduction of new gauge bosons~\cite{xm},
\item four-generation leptons, mixing mainly with the third
generation~\cite{ss,rs},
\item scalar particles which interfere destructively with
the $W$-exchange amplitude in the $\tau$ decay~\cite{f}.
\end{enumerate}
Of course, each of the above possibilities, and their variants, have their
own difficulties. The first one implies a drastic modification of the
Standard Model; the second implies also a modification of the quark
sector, necessary in order to avoid anomalies; finally,
the third one needs a larger Higgs sector (two triplets) in the
Standard Model.

On the other hand, the simplest solution has not been even mentioned
in the literature. It is possible, if the neutrinos are massive, that
a mixing similar to the Kobayashi-Maskawa one~\cite{km} occurs with
three lepton generations. Here we will consider the analysis of the
experimental data with five free parameters, three angles and two
neutrino mass differences from Ref.~\cite{bk}. The effects of the
Kobayashi-Maskawa mixing in the leptonic sector were considered in
Ref.~\cite{ro}. In particular the effects of such a mixing for the
case of leptonic decays of the $\tau$-lepton were explicitly
considered in Ref.~\cite{ssi}. Here we will consider this scenario in
the context of a possible ``$\tau$ decay puzzle''.

The Kobayashi-Maskawa lepton mixing matrix is
\begin{equation}
\left(\begin{array}{c}
\nu'_e \\ \nu'_\mu \\ \nu'_\tau
\end{array}\right)=\left(
\begin{array}{ccc}
V_{ee}    & V_{e\mu}    & V_{e\tau} \\
V_{\mu e} & V_{\mu\mu}  & V_{\mu\tau} \\
V_{\tau e}& V_{\tau\mu} &V_{\tau\tau}
\end{array}\right)
\left(
\begin{array}{c}\nu_e\\ \nu_\mu \\ \nu_\tau\end{array}\right).
\label{4}
\end{equation}
The unprimed fields are mass eigenstates and
we will not consider the hierarchical mixing~\cite{kn}. In
Ref.~\cite{bk} the Maiani parameterization~\cite{ma} of the mixing
matrix was chosen and two solutions were found for the oscillation
parameters which imply the following ranges for the
diagonal matrix elements $(V_{ee},V_{\mu\mu},V_{\tau\tau})$
\begin{description}
\item[ solution a)]
$1.00\mbox{---}0.98,1.00\mbox{---}0.99,1.00\mbox{---}0.98$
\item[ solution b)]
$1.00\mbox{---}0.97,1.00\mbox{---}0.98,1.00\mbox{---}0.98$
\end{description}
In the theoretical predictions of the $\tau$ partial widths we will
neglect neutrino masses. In fact the neutrino mass is only important
if $m_{\nu_\tau}>100\,\mbox{MeV}$ for $\tau\to
e\nu_\tau\bar\nu_e$-decay, or if $m_{\nu_\tau}>50\,\mbox{MeV}$ for
$\tau\to\mu\nu_\tau\bar\nu_\mu$ decay~\cite{ssi} while the present
current limit is $m_{\tau_\tau}<35\mbox{MeV}$~\cite{pdg}.
We use the following notation:
$\bar{B}^\tau_i={\bar\Gamma}^\tau_i/\Gamma^\tau_{tot}$,
${\bar\Gamma}^\tau_i$ being the $\tau$ partial decay width
into the $i$ charged particle $(e^-,\mu^-,\pi^-,K^-)$ considering the
mixing and $\Gamma^\tau_{tot}$ the total width.

The widths with this kind of mixing are given by
\begin{equation}
{\bar\Gamma}^\tau_l=\frac{\vert V_{\tau\tau}\vert^2\vert V_{ll}\vert^2}
{\vert V_{\mu\mu}\vert^2\vert V_{ee}\vert^2}\Gamma^\tau_l,
\label{5}
\end{equation}
and
\begin{equation}
{\bar\Gamma}^\tau_h=\frac{\vert V_{\tau\tau}\vert^2}
{\vert V_{\mu\mu}\vert^2\vert V_{ee}\vert^2}\Gamma^\tau_h,
\label{6}
\end{equation}
where $l=e,\mu$ and $h=\pi,K$ and $\Gamma^\tau_{l,h}$ are the $\tau$
partial widths without mixing which have appeared in the
literature~\cite{m}. In Eqs.~(\ref{5}) and (\ref{6}) the
denominator comes from the definition of the $G_\mu$ constant in the
$\mu$ decay. Numerically we will consider the hadronic
partial decay width as $\Gamma^\tau_h=\Gamma^\tau_\pi+\Gamma^\tau_K$.

Let us start by the theoretical result for the partial width from
Refs.~\cite{m,ms}, which includes radiative corrections, using the
current data from Ref.~\cite{pdg} and presented with the range
implied by the solution a) above:
\begin{eqnarray}
\bar\Gamma(\tau\rightarrow e^-\nu_\tau\bar\nu_e)&=&
(3.95^{+0.03}_{-0.04}
\mbox{---}4.19^{+0.03}_{-0.04}) \times10^{-13}\mbox{GeV},\\
\bar\Gamma(\tau\rightarrow \mu^-\nu_\tau\bar\nu_\mu)&=&
(3.84^{+0.03}_{-0.04}
\mbox{---}4.17^{+0.03}_{-0.04})\times10^{-13}\mbox{GeV},\\
\bar\Gamma(\tau\rightarrow \pi\nu_\tau)&=&
(2.45\pm0.06\mbox{---}2.71\pm0.07)\times
10^{-13}\mbox{GeV},\\
\bar\Gamma(\tau\rightarrow
K\nu_\tau)&=&(1.60\pm0.04\mbox{---}1.76\pm0.04)\times10^{-14}\mbox{GeV},
\\ \bar\Gamma(\tau\rightarrow
h\nu_\tau)&=&(2.61\pm0.06\mbox{---}2.88\pm0.07)\times10^{-13}
\mbox{GeV}.
\label{7}
\end{eqnarray}
With this mixing we have instead of Eq.~(\ref{2})
\begin{equation}
\left(\frac{G_\tau}{G_\mu}\right)^2=\frac{\vert V_{\mu\mu}\vert^2}
{\vert V_{\tau\tau}\vert^2}
\left(\frac{\tau_\mu}{\tau_\tau}\right)
\left(\frac{m_\mu}{m_\tau}\right)^5B.R.(\tau\to e\bar\nu\nu),
\label{8}
\end{equation}
and using the current $\tau$ lifetime and mass~\cite{d}
\begin{equation}
m_\tau=1776.9\pm0.4\pm0.3,\,\mbox{MeV}\quad
\tau_\tau=(3.00\pm0.05)\times10^{-13}s.
\label{9}
\end{equation}
\begin{equation}
\left(\frac{G_\tau}{G_\mu}\right)^2=\frac{\vert V_{\mu\mu}\vert^2}
{\vert V_{\tau\tau}\vert^2}\times \left\{
\begin{array}{c}
0.941\pm0.024\;\mbox{(world average)}\\
0.967\pm0.018\;\mbox{(BEPC)}.\end{array}\right.
\label{10}
\end{equation}
With the values of the diagonal matrix elements given in solution a)
we obtain  $\vert V_{\mu\mu}\vert^2/\vert
V_{ee}\vert^2=0.98\mbox{---}1.04$. We see that there is consistency
with the value $G_\tau/G_\mu=0.975\pm0.010$~\cite{d}. Eqs.(6)-(10)
are also compatible with the respective experimental branching
ratios. Similar results arise using the matrix elements given in (b).

We can also verify that ratios of partial widths are consistent with
leptonic mixing
 \begin{equation}
\frac{\bar\Gamma^\tau_\mu}{\bar\Gamma^\tau_e}=
\frac{\vert V_{\mu\mu}\vert^2}{\vert V_{ee}\vert^2}
\frac{\Gamma^\tau_\mu}{\Gamma^\tau_e}=0.95\pm0.01\mbox{---}
1.01^{+0.01}_{-0.02}\,,
\label{11}
\end{equation}

\begin{equation}
\frac{\bar\Gamma^\tau_h}{\bar\Gamma^\tau_\mu}=
\frac{1}{\vert V_{\mu\mu}\vert^2}
\frac{\Gamma^\tau_\mu}{\Gamma^\tau_e}=0.68\pm0.02\mbox{---}0.69\pm0.02,
\label{12}
\end{equation}

\begin{equation}
\frac{\bar\Gamma^\tau_h}{\bar\Gamma^\tau_e}=
\frac{1}{\vert V_{ee}\vert^2}\frac{\Gamma^\tau_h}{\Gamma^\tau_e}
=0.66\pm0.02\mbox{---}0.69\pm0.02.
\label{13}
\end{equation}
As before, we show in Eqs.~(\ref{11})-(\ref{13}) the range of the
ratios of the partial widths taking into account the range of the
matrix elements. Again it is possible to verify that there is
consistency with experimental data.

We have shown in this comment that a possible deviation from
$\mu-\tau$ universality if confirmed by future experiments is
sufficiently small to be accounted by three generations mixing in the
leptonic sector.

\acknowledgments
We are very gratefull to Funda\c c\~ao de Amparo \`a Pesquisa do Estado de
S\~ao Paulo (FAPESP) (R.Z.), Coordenadoria de Aperfei\c coamento de Pessoal
N\'\i vel Superior (CAPES) (O.L.G.P.) for full financial support and
Con\-se\-lho Na\-cio\-nal de De\-sen\-vol\-vi\-men\-to Cien\-t\'\i
\-fi\-co e Tec\-no\-l\'o\-gi\-co (CNPq) (V.P.) for
partial financial support.

\end{document}